\documentclass[a4paper]{jpconf}
\usepackage{graphicx}
\usepackage{iopams}
\begin{document}
\title{Thermoelectric effects in correlated quantum dots and molecules}

\author{V. Sch\"{o}ps$^{1,2}$, V. Zlati\'c$^{1,3}$ and T. A. Costi$^1$}

\address{$^1$ Institute of Solid State Research, Research Centre J\"{u}lich, 
52425 J\"ulich, Germany}
\address{$^2$ Department of Physics, University of Konstanz, 78457 Konstanz, Germany}
\address{$^3$ Institute of Physics,  10001 Zagreb, Croatia}

\ead{t.costi@fz-juelich.de}

\begin{abstract}
We investigate thermoelectric properties of correlated quantum dots and molecules, 
described by a single level Anderson model coupled to conduction electron leads, by
using Wilson's numerical renormalization group method. 
In the Kondo regime, the thermopower, $S(T)$, 
exhibits two sign changes, at temperatures $T=T_{1}$ and 
$T=T_{2}>T_{1}$. We find that $T_{2}$ is of order the level width $\Gamma$
and $T_{1}> T_{p}\approx T_{K}$, where $T_{p}$ is the position of the Kondo 
induced peak in the thermopower and $T_{K}$ is the Kondo scale. No sign 
change is found outside the Kondo regime, or, for weak correlations, 
making a sign change in $S(T)$ a particularly 
sensitive signature of strong correlations and Kondo physics.
For molecules, we investigate the effect of screening by conduction
electrons on the thermoelectric transport. We find that a large screening
interaction enhances the figure of merit in the Kondo and mixed valence regimes. 
\end{abstract}

\section{Introduction}
Materials with high thermoelectric efficiency are currently under intense 
theoretical and  experimental investigation, largely due to the prospect 
of applications, e.g., for the conversion of waste heat into electricity, or 
for refrigeration and on-chip cooling in microelectronics 
\cite{mahan.98,kanatzidis.10}. Apart from applications, thermoelectric 
materials also serve as an interesting testing ground for new theoretical 
approaches to thermoelectric transport in solids\cite{zlatic.94,costi.94,zlatic.05}. 
Coulomb correlations and the Kondo effect give rise to 
large thermoelectric coefficients in strongly correlated bulk materials 
\cite{arita.08,lackner.06,bentien.07,sales.96}.
They are also important in nanoscale semiconducting devices and in 
molecular transistors, posing interesting theoretical and experimental 
challenges for the understanding of electrical and thermal transport 
in such systems\cite{scheibner.05,kim.02,dong.02,roch.09,parks.10,koch.04,reddy.07}.  

In this paper we address the thermoelectric properties of two related
systems: a nanoscale size quantum dot exhibiting the Kondo effect, which we 
describe in terms of a single level Anderson impurity model with two conduction
electron leads at fixed chemical potentials, and, a molecular transistor described
by an extension of the above model to include a local screening of the Coulomb
interaction on the molecule by the electrons in the leads. The quantum dots or
molecules that we consider can be tuned from the Kondo to the mixed 
valence and empty orbital regimes by a gate voltage
\cite{goldhaber-gordon.98,cronenwett.98}. 
Very recent experiments on nanoscale quantum dots 
\cite{scheibner.05} are beginning to probe the effect of Kondo correlations on 
the thermopower. 

We use Wilson's numerical renormalization group (NRG) method
\cite{wilson.75,bulla.08} to calculate the linear conductance, thermopower
and thermal conductance as a function of temperature and gate voltage. The technique
gives reliable results for transport properties in all regimes of 
interest\cite{costi.94}. The recently introduced full density matrix (FDM) 
approach \cite{fdm.07}(see also \cite{peters.06}) is used. 
This allows calculations of dynamical properties at all 
excitation energies $\omega$ relative to the temperature $T$, thereby 
simplifying the calculation of transport properties which require knowledge 
of excitations, $\omega$, above and below the temperature.

\begin{figure}[t]
\includegraphics[width=10cm]{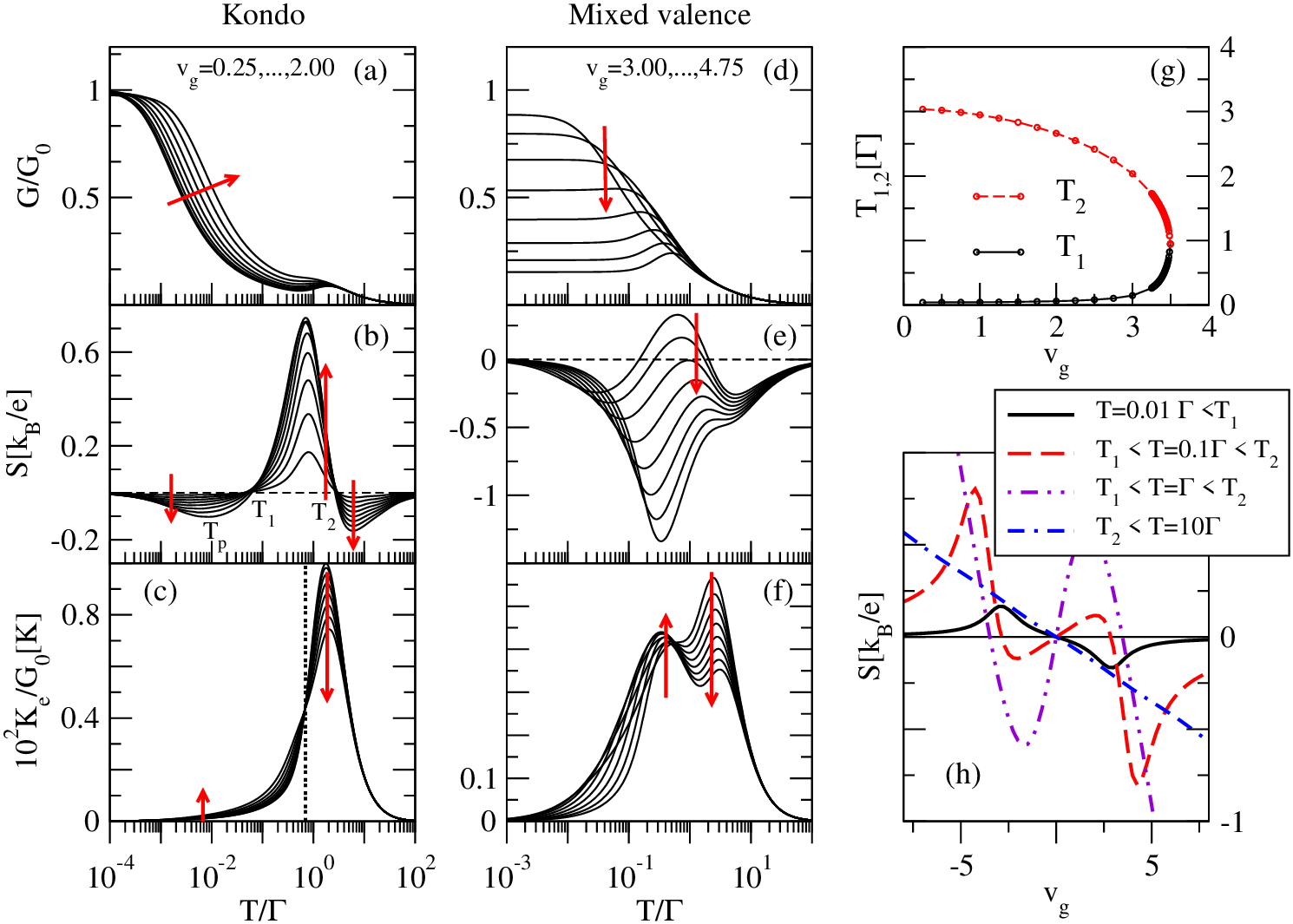}
\hspace{2pc}
\begin{minipage}[b]{14pc}
\caption{\label{figure1}
Conductance, $G$, thermopower, $S$, 
and, thermal conductance, $K_{\rm e}$, versus $T/\Gamma$, for $U=4\Gamma=0.08$ 
and gate voltages, ${\rm v}_g=(\varepsilon_{d}+U/2)/\Gamma$, in the Kondo 
(a-c) and mixed valence (d-f) regimes ($G_{0}=2e^{2}/h$). The ${\rm v}_g$ are indicated in
top panels and increase by $0.25$. Arrows show trends with increasing ${\rm v}_g$.
$T_{p}$: position of low temperature peak in $S$. Temperatures $T_{1,2}$ at which $S(T)$ 
changes sign are shown in (g) versus ${\rm v}_{g}$. (h): gate voltage dependence of $S$ 
at representative temperatures $T$ relative to $T_{1}(0)$
and $T_{2}(0)$.
}
\end{minipage}
\end{figure}

\section{Model and transport properties}
\label{model}
We describe both the nanoscale quantum dot and the molecular transistor by 
the single level Anderson impurity model with two conduction electron leads
\begin{eqnarray}
H &=& \sum_{\alpha k\sigma}\epsilon_{\alpha k\sigma}c_{\alpha k\sigma}^{\dagger}c_{\alpha k\sigma} 
+ \sum_{\sigma}\varepsilon_{d}\,d_{\sigma}^{\dagger}d_{\sigma} + 
U n_{d\uparrow}n_{d\downarrow} + H_{\rm screening}
 + \sum_{\alpha k\sigma}t_{\alpha}(c_{\alpha k\sigma}^{\dagger}d_{\sigma}+h.c.).
\label{qdot-two-leads}
\end{eqnarray}
Here, $\epsilon_{\alpha k\sigma}$ is the kinetic energy of 
electrons with wavenumber $k$ and spin $\sigma$ in lead $\alpha=(L,R)$, 
$\varepsilon_{d}$ is the local level energy, $U$ is the local Coulomb repulsion 
and $t_{\alpha}$ is the tunnel matrix element
of the local level to conduction electron states in lead $\alpha=(L,R)$. 
$H_{\rm screening} = U_{dc}(n_{d}-1)(n_{0}-1)$, with $n_{d}$ the local level
occupancy and $n_{0}$ the local occupancy of the lead electrons, represents the
screening interaction for the case of the molecule. 
We assume symmetric couplings of the dot to
both leads and use the full width at half maximum, $\Gamma=0.02$, as the energy unit
throughout.

Thermoelectric transport 
is calculated for a situation in which a small external bias voltage, 
$\delta V=V_{L}-V_{R}$, and a small temperature gradient $\delta T$
is applied between left and right leads
following \cite{kim.02,dong.02}. To linear order,
the following expressions for the electrical conductance, $G(T)$, the thermal 
conductance, $K_{\rm e}(T)$, and the thermoelectric power, $S(T)$,
are obtained
\begin{eqnarray}
 G(T) &=& e^{2}I_{0}(T)\label{transport-expressions-G}\\
S(T) &=& -\frac{1}{|e|T}\frac{I_{1}(T)}{I_{0}(T)}\label{transport-expressions-S}\\
K_{\rm e}(T) &=& \frac{1}{T}\left[I_{2}(T) - \frac{I_{1}^{2}(T)}{I_{0}(T)} \right]
\label{transport-expressions-K}
\end{eqnarray}
where $I_{n},n=0,1,2$ are the transport integrals
\begin{equation}
I_{n}(T) = \frac{2}{h}\int d\omega\; \omega^{n}{\cal T}(\omega)(-\frac{\partial f}{\partial \omega}).
\label{trans-dot}
\end{equation}
Here, $e$ denotes the magnitude of the electronic charge and $h$ denotes Planck's constant.
The  transmission function ${\cal T}(\omega)=\pi{\Gamma}A(\omega)$, where $A(\omega)$ is
the d-level spectral density.

\section{Results for quantum dots: $U_{dc}=0$}
The temperature dependence of the electrical conductance, $G(T)$, thermopower, $S(T)$, and
electronic part of the thermal conductance, $K_{\rm e}(T)$, are shown in 
Fig.~\ref{figure1}a-c in the Kondo regime and in Fig.~\ref{figure1}d-f in the mixed valence regime
\cite{costi.10}. The thermopower exhibits two sign changes in the Kondo regime at $T=T_{1}({\rm v}_{g})$
and $T=T_{2}({\rm v}_{g})$ and no sign changes on entering the mixed valence regime (see Fig.~\ref{figure1}e).
The gate voltage dependence of $S$ in Fig.~\ref{figure1}h shows three characteristic types 
of behaviour depending on the $T$ relative to $T_{1}(0)$ as explained in \cite{costi.10}.

\section{Results for molecules: $U_{dc}\ge 0$}
For molecules in metallic break junctions, we expect additional effects
to be important for thermoelectric properties, e.g. local phonon modes 
and screening by the conduction electrons in the leads. Here, we investigate the effect of
the latter, and note that the effect of a strong local electron-phonon coupling 
can be approximately simulated by choosing a large screening interaction $U_{dc}>U$.

\begin{figure}[t]
\includegraphics[width=10cm]{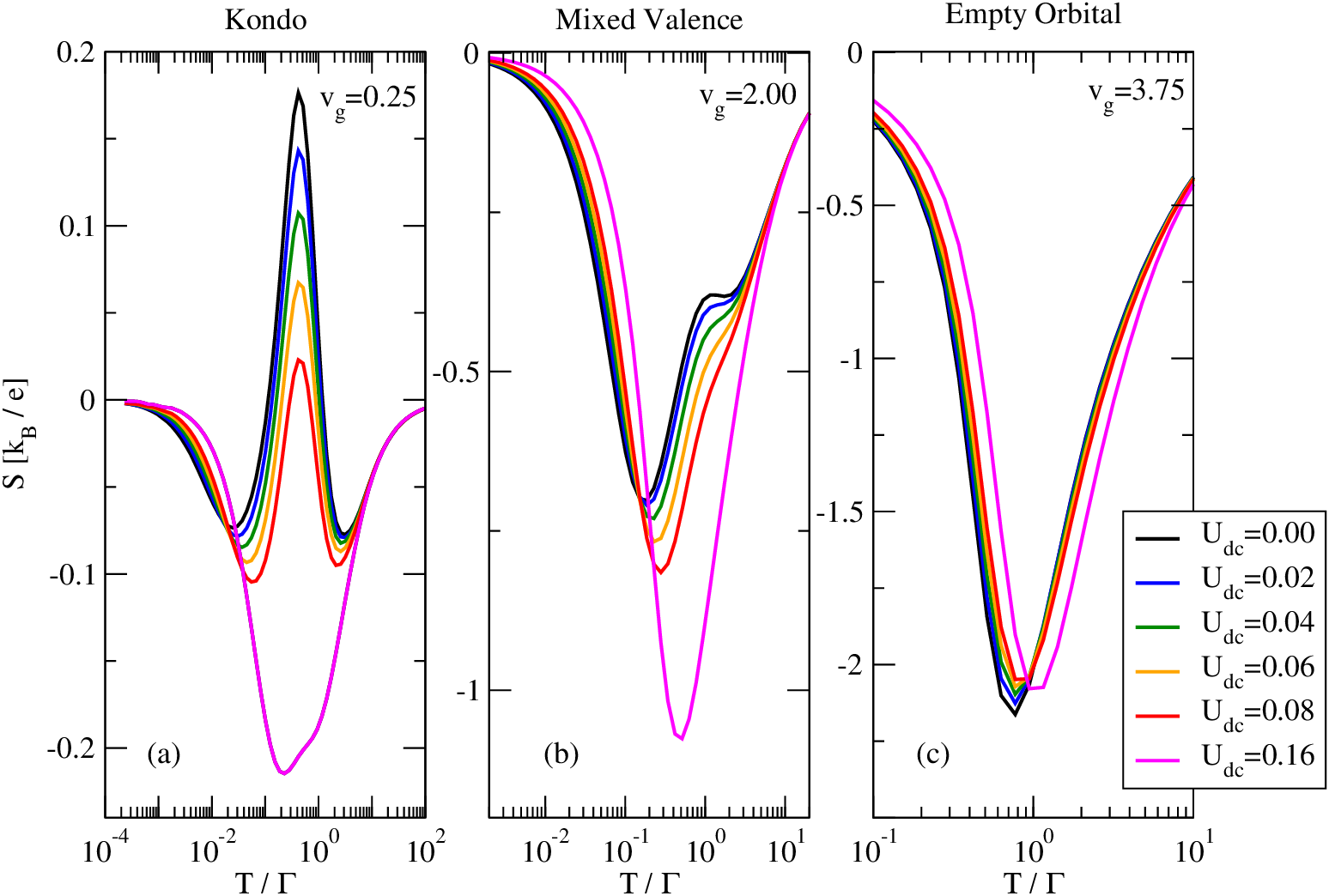}
\hspace{2pc}
\begin{minipage}[b]{14pc}
\caption{\label{figure2}
Temperature dependence of the thermopower, $S(T)$, 
for $U=4\Gamma=0.08$ in the Kondo, (a), mixed valence, (b), and empty orbital, (c), 
regimes for increasing values of the screening interaction $U_{dc}$. The case of
large screening $U_{dc}=0.16>U$ is qualitatively similar to a large local electron-phonon
coupling and results in destruction of Kondo correlations
at finite gate voltages. Screening has a negligible effect in the empty orbital
regime, but affects significantly $S(T)$ in the other regimes.
}
\end{minipage}
\end{figure}

\begin{figure}[t]
\includegraphics[width=10cm]{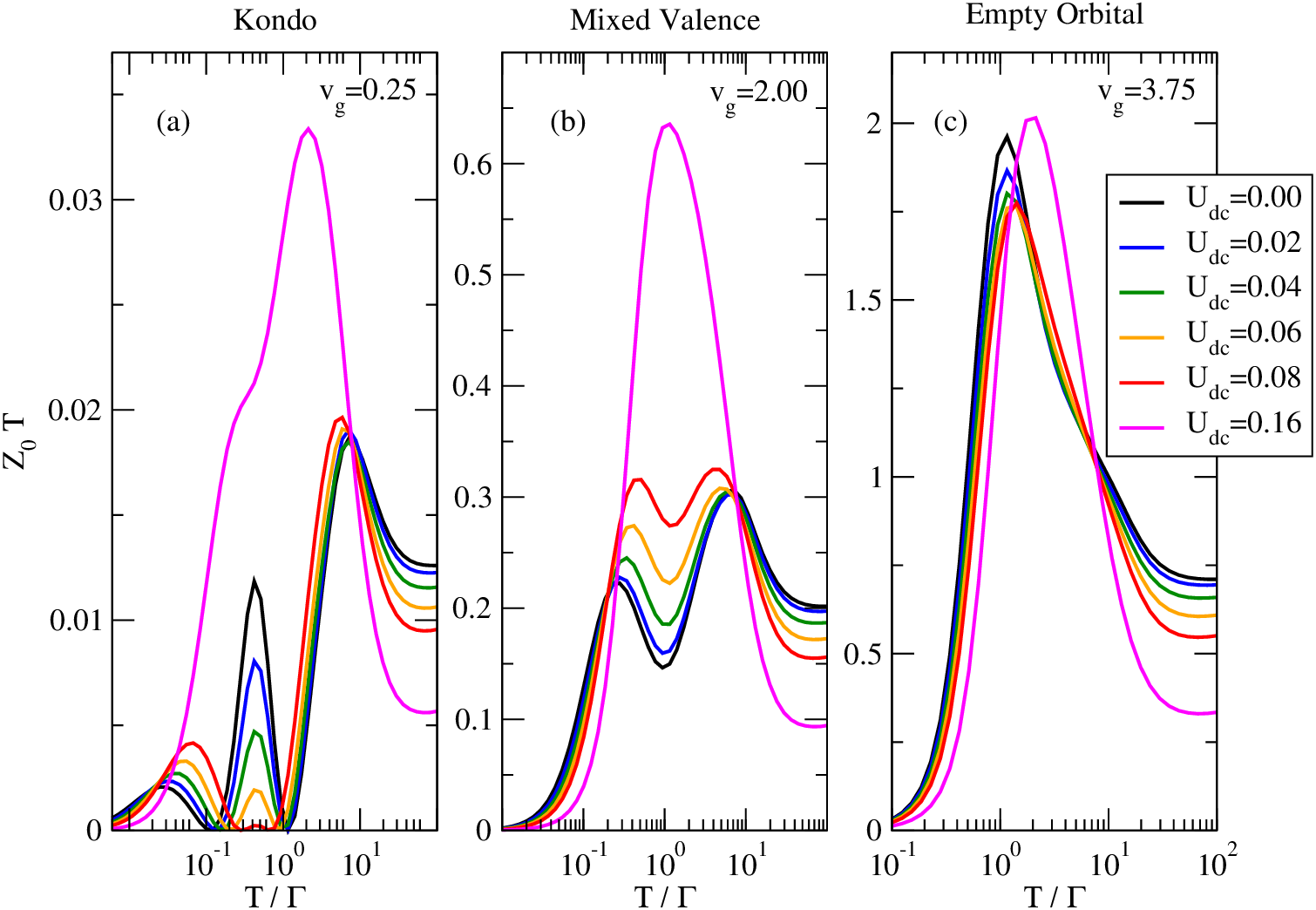}
\hspace{2pc}
\begin{minipage}[b]{14pc}
\caption{\label{figure3}
Temperature dependence of the ``figure of merit'' $Z_{0}T=GS^{2}T/K_{\rm e}$ for
 $U=4\Gamma=0.08$ and increasing values of the screening interaction $U_{dc}$ in 
different regimes (by reference to the $U_{dc}=0$ case). The true figure of merit
$ZT=GS^{2}T/(K_{\rm e}+K_{\rm ph})$ includes a material specific phonon contribution, 
$K_{\rm ph}$, which we have neglected, but needs to be included whenever phonons 
dominate thermal transport. Provided all scales are less than the Debye temperature $\theta_{D}$,
we expect that $ZT\approx Z_{0}T$.
}
\end{minipage}
\end{figure}

For moderate screening interactions $U_{dc}<U$, we recover the expected 
downward renormalization of the local Coulomb repulsion $U\rightarrow U^{*}<U$ and an
upward excitonic renormalization of the effective hybridization with increasing 
$U_{dc}$ leading to a weakening of Kondo correlations.
Correspondingly, transport properties are significantly affected in the
Kondo and mixed valence regimes, as shown in Fig.~\ref{figure2} for the thermopower and
in Fig.~\ref{figure3} for the figure of merit. For large screening interactions $U_{dc}>U$ 
we find a destruction of Kondo correlations for gate voltages  ${\rm v}_{g}>0$ away from
particle-hole symmetry with an enhancement of the figure of merit for ``Kondo'' and ``mixed
valence'' regimes (referred to the $U_{dc}=0$ case).

\section{Conclusions}
The thermoelectric power of, (i), quantum dots, and, (ii), molecules in break junctions,
has been investigated using a single-level Anderson model. We showed that the thermopower
exhibits sign changes as a function of temperature and gate voltage. These can be 
used as sensitive probes of Kondo correlations. Recent measurements on Kondo correlated
quantum dots \cite{scheibner.05} are in qualitative agreement with our model calculations
\cite{costi.10}. The effect of lead electron screening on molecular transport is shown
to be significant in the Kondo and mixed valence regimes, where we find that an enhanced
figure of merit can result. In future, it would be interesting to include a local 
electron-phonon coupling of the molecular state in order to investigate the
effect of this interaction on molecular transport in detail\cite{leijnse.10}.

\ack
Support from the Forschungszentrum J\"{u}lich (V.Z.), the SFB 767 (V.S.) and 
supercomputing support from the John von Neumann Institute for Computing (J\"ulich)
is acknowledged.

\end{document}